\documentclass[journal]{IEEEtran}
\usepackage{bm}
\usepackage[cmintegrals]{newtxmath} 

\usepackage{epsfig}
\usepackage{graphicx}
\usepackage{subfigure}
\usepackage{caption}
\usepackage{placeins}
\usepackage{float}
\usepackage{boxedminipage}

\usepackage{color}

\usepackage{cite}
\usepackage{tabularx,colortbl}
\usepackage[ruled,vlined,linesnumbered]{algorithm2e}
\usepackage{amstext}
\usepackage{amsmath}
\usepackage{amsfonts}            
\usepackage[cal=cm,bb=ams,scr=boondoxo] {mathalfa}

\usepackage{amsthm} 
\allowdisplaybreaks[4] 

\usepackage{extarrows}

\usepackage{etoolbox}
\usepackage{mdwlist}
\usepackage{booktabs} 
\makeatletter
\patchcmd{\@maketitle} 
{\addvspace{0.5\baselineskip}\egroup} 
{\addvspace{-1.5\baselineskip}\egroup} 
{}
{}

\newtheorem*{assumption*}{\assumptionnumber}
\providecommand{\assumptionnumber}{}
\newtheorem*{definition*}{\definitionnumber}
\providecommand{\definitionnumber}{}



\usepackage[pagewise,switch,mathlines]{lineno}

\begin{document}
	%
	\title{\huge Multiple Sources Localization with Sparse Recovery\\ under Log-normal Shadow Fading}
	%
	%
	%
	
	\author{
		Yueyan~Chu,~Wenbin~Guo,~\IEEEmembership{Member,~IEEE},~Kangyong~You,~\IEEEmembership{Student~Member,~IEEE,}\\~Lei~Zhao,~Tao~Peng,~and~Wenbo~Wang,~\IEEEmembership{Senior~Member,~IEEE}
		
				
		
			
	}

\maketitle
\begin{abstract}
Localization based on received signal strength (RSS) has drawn great interest in the wireless sensor network (WSN). In this paper, we investigate the RSS-based multi-sources localization problem with unknown transmitted power under shadow fading. The log-normal shadowing effect is approximated through Fenton-Wilkinson (F-W) method and maximum likelihood estimation is adopted to optimize the RSS-based multiple sources localization problem. Moreover, we exploit a sparse recovery and weighted average of candidates (SR-WAC) based method to set up an initiation, which can efficiently approach a superior local optimal solution. It is shown from the simulation results that the proposed method has a much higher localization accuracy and outperforms the other.
\end{abstract}


\begin{IEEEkeywords}
	Received signal strength, multi-sources localization, shadow fading, maximum likelihood estimation, sparse recovery and weighted average of candidates.
\end{IEEEkeywords}

%
\IEEEpeerreviewmaketitle

\section{Introduction}
%
%
%
%

\IEEEPARstart{L}{ocalization} has been playing an increasingly essential role in both military and commercial applications\cite{Saeed2019State}. Positioning systems are divided into active positioning and passive positioning according to whether the reconnaissance equipment emits radio signals. Active positioning is charactered by all-weather and high-speed positioning, which is widely used in radio-frequency identification (RFID)\cite{Zhi2010RFID}, traffic alert and collision avoidance systems (TCAS)\cite{Williamson1989TCAS} and other applications. Unlike active positioning, where radio signals are emitted by transmitters, passive positioning has received considerable attention due to good concealment, strong anti-interference ability and low cost in recent years, across from Global Positioning System (GPS), wireless sensor network (WSN), tracking radars to vision system\cite{Zekavat2019Handbook}. In particular, localization in WSNs have a broad spectrum of applications in fields such as disaster monitoring, smart home, indoor navigation and animal tracking\cite{Kishore2016MANET,Ransing2015Smart,Zafari2019Survey,Vera2019Design}.

The main method of obtaining the target position in WSNs is to exploit the measurement of the signal received by sensors employed in the region of interest (ROI). The most commonly used measurements are time of arrival (TOA)\cite{Alsindi2009Measurement,Wang2020Multipath}, time difference of arrival (TDOA)\cite{Qiao2014Nonlinear,Meng2016Optimal}, direction of arrival (DOA)\cite{Stein2017CaSCADE,Reddy2015Reduced}, angle of arrival (AOA)\cite{Inserra2013Frequency,Steendam2018Positioning} and received signal strength (RSS)\cite{Liu2016RSS,You2020Parametric,Abeywickrama2018Wireless}. Precise timing synchronization of each sensor node (SN) is required for TOA and TDOA, while localization for DOA and AOA need each SN to provide a directional antenna array. These methods have high accuracy, however, they are difficult to arrange in low-cost WSNs. Compared with the above methods, which have strict requirements for hardware equipment, RSS-based localization becomes appealing due to its simplicity and low energy-consumption, and is easily obtained from existing electronic systems. Moreover, it can be fused into other systems as a hybrid localization approach\cite{Katwe2020NLOS,Coluccia2018On,Catovic2004The,Tomic2017RSS/AOA,Tomic2016Closed,Xu2020Optimal,Wang2013Cramer}. Therefore, RSS-based localization has aroused extensive attention in WSNs. 

\subsection{Related Works}
In the following, we review research works related to RSS-based localization including the single-source localization and multi-source localization, where methods based on optimization, compressed sensing (CS)\cite{You2020Parametric,You2018Grid} and machine learning\cite{Prasad2018Machine} are proposed, respectively. Especially, range-based solution are more popular than range-free approach according to the path loss model for its simplicity.

\subsubsection{Single-source Localization}
Early method for RSS-based single-source localization (RSSL) is to determine the location of the source by trilateration, which converts the RSS value into distance between the source and SNs based on a certain propagation model\cite{Manolakis1996Efficient}. As a result of vulnerability against noise, propagation model and sensors, trilateration method is replaced by methods based on optimization, which considers unknown propagation parameters and noise. A least-square (LS) algorithm is proposed in \cite{Lee2009Location}, where unknown transmit power is eliminated through energy ratios between sensors. A maximum likelihood estimation (MLE) method is put forward in \cite{Patwari2003Relative}, which enhances the robustness against noise. Since the RSSL is a highly non-convex and non-linear parameter estimation problem and the MLE method has a high computational complexity in order to find a global optimal solution, methods by applying efficient convex relaxations that are based on second-order cone programming (SOCP) and semidefinite programming (SDP) to relax the MLE are proposed in \cite{Tomic2013RSS,Tomic2015RSS}. Besides, the RSSL is studied in \cite{Vaghefi2013Cooperative,Gholami2013RSS,Sari2018RSS} when the propagation parameters are unknown.

\subsubsection{Multi-source Localization}
Gradually, RSS-based multi-source localization (RMSL) has extensively aroused interest of researchers. RMSL, more specifically, is RSS-based co-channel multi-source localization, where the RSS measurement of a SN is the linear superposition of multiple source powers and there is no longer a clear correspondence between RSS and sensor-source distance, which makes RMSL more challenging. A MLE-based method for RMSL is firstly proposed in \cite{Sheng2005Maximum}, which considers a multiresolution search algorithm and an expectation-maximization (EM) like iterative algorithm to start a coordinate search in the ROI discreted into a set of grid points (GPs). Additionally, distributed particle filters with Gaussian mixer approximation are proposed to localize and track multiple moving targets in \cite{Sheng2005Distributed}. However, these methods brings high computational complexity. Instead, an approach based on CS is proposed for RMSL \cite{Cevher2008Distributed,You2018Grid,You2020Parametric}, which utilizes the sparsity of the source locations in the grid space. As a result, the estimated locations of the sources on the candidate GPs can be obtained via general sparse recovery algorithms.

\subsubsection{RSS-based Localization under Shadow Fading}
It is a remarkable fact that the aforementioned works in the RMSL mainly consider the obstacle-free environment. While, shadow fading, belonging to multiplicative noise, is caused by obstructions in an environment with obstacles, which exert a tremendous influence on RSS-based localization. Numerous studies show that the shadow fading channel is widely used to establish a radio propagation model in wireless communication applications such as cellular communication, surface communication, and broadcast reception. Hence, considerable RSS-based localization approaches under shadow fading have been proposed in the past few years. A SDP-based method is proposed in \cite{Vaghefi2013Cooperative}. Besides, a MLE-based method is presented to solve the problem of RSSL under mixture shadow fading\cite{Kurt2017RSS}. The multiplicative noise is converted into additive noise by expressing the RSS measurement in the logarithmic domain in the approaches mentioned below for RSSL to realize the localization.

Nevertheless, it is a huge challenge for RMSL under shadow fading. The RSS value of each SN is a linear superposition of multiple signals from sources, that is, the sum of LN random variables in terms of probability. Hence, the signal and noise cannot be separated in the logarithmic domain. Furthermore, since moment generating function (MGF) of LN random variables is not defined\cite{Heyde1963property}, the probability density function (PDF) of log-normal sum (SLN) distributions cannot be expressed exactly\cite{Fenton1960The}. Consequently, it is impossible to propose algorithms based on classical estimation such as MLE, minimum variance unbiased (MVU) estimation, Bayesian estimation, etc. To the best of our knowledge, a minimum mean square error (MMSE) based method, which is an unbiased estimation, is proposed in \cite{Zandi2018RSS,Zandi2019Multi} for RMSL in recent years, which approximates SLN via LN.

\subsection{Contributions}
In this paper, we investigate the RMSL with unknown transmitted power in a log-normal shadow fading environment. The closed-form expression of SLN distributions is shown motivated by \cite{Zandi2018RSS,Zandi2019Multi}. Then, a MLE-based method based on the probability distribution is put forward and a proper initial-point-determination approach is utilized. The main contributions of this paper are illustrated as follows:
\begin{itemize}
	\item
	Different from the MMSE method in \cite{Zandi2018RSS,Zandi2019Multi}, a maximum likelihood estimation-based algorithm in terms of probability is proposed to formulate an optimization problem, which can be solved by a gradient descent method. Then, we make use of the Gradient Projection (GP) method to solve this non-convex problem and obtain the locations of the sources. 
	\item
	There are multiple local optimal solutions due to the non-linearity and non-convexity of the aforementioned optimization problem, which is hard to deal with. Therefore, a novel inital-point-determination method based on sparse recovery and weighted average of candidates (SR-WAC) is proposed to approach a better local optimal solution against the problem that the gradient descent method is sensitive to the initial value.
\end{itemize}

The remainder of this paper is organized as follows. Section II presents a system model for RMSL. F-W method is exploited to approximate SLN by LN and a MLE-based method is derived in Section III. The proposed SR-WAC algorithm for initial-point-determination is given in Section IV. Numerical simulation results are shown to evaluate the performance of the proposed algorithm in Section V. Section VI concludes our paper.

{\it Notation:} $\mathbb{N}$ denotes the set of positive integers. $x_i$ is the $i$th element of a vector $\bm{x}$. $[\bm{A}]_{i, j}$ is the (i, j)th element of a matrix $\bm{A}$. $\left\|\cdot\right\|_0$ and $\left\|\cdot\right\|_0$ are $l_0$-norm and $l_1$-norm, respectively. $\bm{I}$ is the identity matrix. $(\cdot)^T$ and $(\cdot)^{-1}$ denote transpose and inverse operators. $\mathbb{E}\left\{\cdot\right\}$ and $\mathbb{D}\left\{\cdot\right\}$ denote the mean and variance, respectively. $\bm{1}_{N}$ is the all ones vector of dimension $N$. $P_r(\cdot)$ denotes the probability and $std(\cdot)$ is the standard deviation.

\section{System Model}

\begin{figure}[t]	
	\centering
	\epsfig{figure=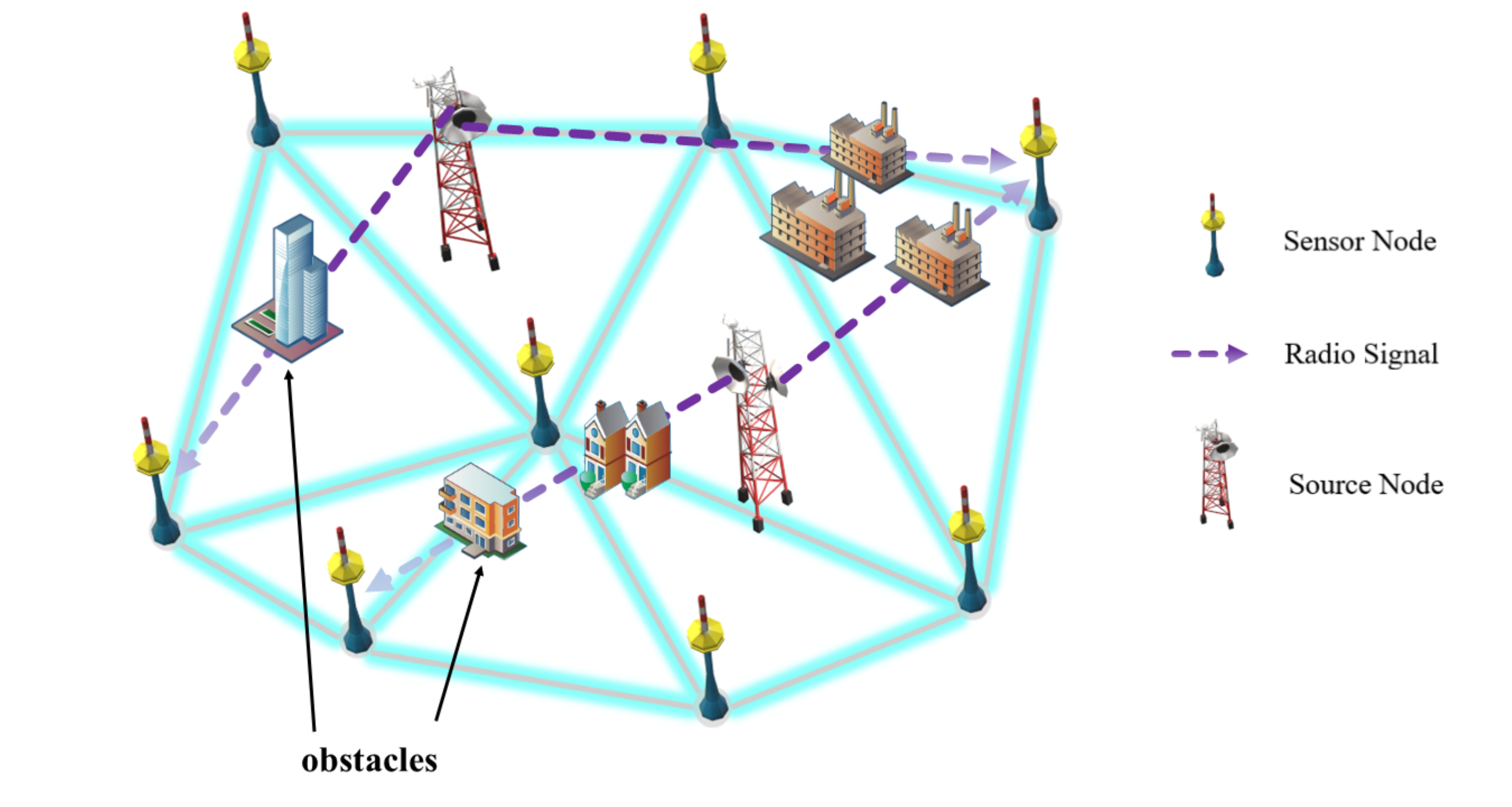,width=8cm}
	\caption{System model for RMSL under shadow fading}
	\label{fig_Sysmodel}
\end{figure}

In this section, we present the system model in a log-normal shadowing fading environment with unknown transmitted power. 
A two-dimensional (2-D) rectangular area with length and width respectively being $l$ and $w$, is considered, which consists of $M\in\mathbb{N}$ sensor nodes (SNs) and $K\in\mathbb{N}$ target sources (TSs), where $K\geq2$. Let $\bm{t}_k=[u_k,v_k]^{T}$ and $\bm{a}_m=[u_m,v_m]^{T}$ be the locations of the TS and the SN, respectively, where $k=1,2,\dots,K$ and $m=1,2,\dots,M$. Sources and sensors are randomly distributed in the ROI as shown in Fig. \ref{fig_Sysmodel}.

Firstly, the case of the single source is discussed and the Euclidean distance $d_m$ between the TS and the $m$th SN can be denoted as:
\begin{equation}\label{Euclidean_distance}
	d_m = \left\|\bm{t}-\bm{a}_m\right\|_2 = \sqrt{(u-u_m)^2+(v-v_m)^2}
\end{equation}
where $\bm{t}=[u,v]^T$ represents the two-dimensional coordinate of the TS. A poor performance of the localization is observed under shadow fading caused by obstacles. Hence, the received power is commonly modeled by the log-normal shadow fading. We assume that the TS transmits a signal and SNs can obtain the received signal strength (RSS) from the TS. The RSS, $P_m$, at the $m$th SN, is presented by \cite{Bandiera2015Cognitive,Gholami2013RSS}
\begin{equation} \label{RSSL_expression}
	P_m=P_0-10 \alpha lg(\frac{d_m}{d_0})+n_m, \quad m=1,2,\cdots,M
\end{equation}
where $P_0$ (in dBm) is the reference power at distance $d_0$ from the TS, $\alpha$ is the path-loss exponent (PLE), which is 2 in the free space, 2 $\sim$ 2.5 in the open environment, 2.5 $\sim$ 3.0 in the semi-open environment and 3.0 $\sim$ 3.5 in the closed environment. $n_m$ is a zero-mean Gaussian random variable with variance $\sigma^2$, which models the log-normal shadow fading, i.e., $n_{m} \sim \mathcal{N} (0,\sigma^2)$. 

Then, for a scenario of multiple co-channel TSs, the RSS at the $m$th SN are able to be denoted by
\begin{equation}
	r_m=\sum_{k=1}^{K}c_0 P_k d_{mk}^{-\alpha} 10^{\frac{n_{mk}}{10}}+\varepsilon_m,\quad m=1,2,\cdots,M
\end{equation}
where $P_k$ is the transmitted power of the $k$th TS, $P_k \in [P_{low},P_{high}]$. $d_{mk}$ represents the Euclidean distance between the $m$th SN and the $k$th TS. $n_{mk}$ is an independent and identically distributed zero-mean Gaussian random variable with variance $\sigma_{s}^2$, which models the log-normal shadow fading, i.e., $n_{m} \sim \mathcal{N} (0,\sigma_{s}^2)$. $\varepsilon_m$ is the receiver noise modeled as additive white Gaussian noise. The coefficient $c_0$ is given by \cite{Zekavat2019Handbook}
\begin{equation}
	c_0=\frac{G_t \mathcal L_t^{-1} G_r \mathcal L_r^{-1} \lambda^2}{(4\pi)^2}, 
\end{equation}
where $G_t$ and $\mathcal L_t$ (or $G_r$ and $\mathcal L_r$) denote antenna gain and system loss factors of a transmitter (or a receiver), respectively. $\lambda$ is the wavelength of the transmitted signal. In order to simplify the model, we reasonably assume $c_0=1$. Moreover, since shadowing has much bigger effect compared with the white Gaussian noise (WGN), the WGN, i.e., $\varepsilon_m$, is ignored in our model. Therefore, the RSS in our work will be expressed as:
\begin{equation} \label{RMSL_expression}
	r_m=\sum_{k=1}^{K}c_0 P_k d_{mk}^{-\alpha} 10^{\frac{n_{mk}}{10}},\quad m=1,2,\cdots,M
\end{equation}

\section{Maximum Likelihood Estimation Method Based on F-W Approximation}
In this section, many approximation methods are reviewed due to a bad behavior of log-normal sum (SLN) distributions and the F-W method is given to approximate SLN corresponding to our signal model (\ref{RMSL_expression}) by LN as the first step for RMSL. Then, unlike the MMSE algorithm in \cite{Zandi2018RSS,Zandi2019Multi}, we propose a method based on Maximum Likelihood Estimation (MLE)  according to the probability model for RMSL and the optimization problem is formulated.
\subsection{The Review of Approximation Methods}
The RSS is the sum of log-normal (LN) random variables according to (\ref{RSSL_expression}) and (\ref{RMSL_expression}), which is common in communications problems, such as the analysis of co-channel interference in cellular mobile systems, the error performance analysis of an ultra-wide-band (UWB) system \cite{Liu2003Error} and the computation of outage probabilities \cite{Beaulieu2004Optimal}. However, since general closed-form expressions for the log-normal sum (SLN) PDF are unknown, traditional estimator based on maximum likelihood and minimum variance unbiased can not be employed directly.

There are several approximation methods in the literature \cite{Beaulieu1995Estimating} based on LN to approximate SLN, such as Fenton-Wilkinson (F-W) method based on moments matching \cite{Fenton1960The}, Schwartz-Yeh (S-Y) method based on exact moments of two LN random variables via iterations \cite{Schwartz1982On}, Beaulieu-Xie method using a linearizing transform with a linear minimax approximation \cite{Beaulieu2004Optimal} and Schleher's method based on cumulants matching \cite{Schleher1977Generalized}. Compared with other methods, the F-W method is simple without iterations, and tries to directly match the first and second moments of SLN by LN. Most importantly, only the F-W method provides a closed-form expression for all the moments of SLN distributions after approximation. To sum up, we choose the F-W method.

\subsection{The Approximation of SLN Distribution by F-W Method}
We first assume that $r_{mk}$ is a LN random variable, which denotes the RSS at the $m$th SN from the $k$th TS, i.e., $r_{mk}=P_{k}d_{mk}^{-\alpha}10^{\frac{n_{mk}}{10}}$. Then, the RSS at the $m$th SN, $r_{m}=\sum r_{mk}$, can be approximated as
\begin{equation}
	r_{m} \approx \widetilde{r}_{m} = e^{X},
\end{equation}
where $X$ is a normal random variable with mean $\mu_m$ and variance $\sigma_m^2$. Then, the mean and variance of $\widetilde{r}_{m}$ can be obtained by \cite{Fenton1960The}
\begin{subequations} \label{character_LN}
	\begin{gather}
		\mathbb{E}\{\widetilde{r}_{m}\}=e^{\mu_m+\frac{\sigma_m^2}{2}} \label{character_LN1},
		\\	
		\mathbb{D}\{\widetilde{r}_{m}\}=(e^{\sigma_m^{2}}-1)e^{2\mu_m+\sigma_m^2} \label{character_LN2}.
	\end{gather}	
\end{subequations}
Next, according to (\ref{character_LN1}), the $l$th moment of $r_{mk}$ is presented by
\begin{equation} 
	\mathbb{E}\{r_{mk}^{l}\}=(P_{k}d_{mk}^{-\alpha})^{l}\beta^{l^{2}},
\end{equation}
where $\beta=\frac{(\ln10)^{2}\sigma_s^{2}}{200}$, which we assume is known. Thus, the mean and variance of $r_{mk}$ are written respectively as
\begin{subequations} \label{character_rmk}
	\begin{gather}
		\mathbb{E}\{r_{mk}\}=P_{k}d_{mk}^{-\alpha}\beta, 
		\\
		\mathbb{D}\{r_{mk}\}=\mathbb{E}\{r_{mk}^{2}\}-\mathbb{E}^{2}\{r_{mk}\}=(P_{k}d_{mk}^{-\alpha})^{2}\beta^{2}(\beta^{2}-1).
	\end{gather}
\end{subequations}
Let us assume $r_{mk}$ from different TSs are independent, and according to (\ref{character_LN}) and (\ref{character_rmk}), the mean and variance of $\widetilde{r}_{m}$ are derived to match the 1th and 2th moments of SLN by
\begin{subequations} \label{character_rm}
	\begin{gather}
		e^{\mu_{m}+\frac{\sigma_{m}^2}{2}}=\mathbb{E}\{r_{m}\}=\beta\sum_{k=1}^{K}P_{k}d_{mk}^{-\alpha}, 
		\\
		(e^{\sigma_{m}^{2}}-1)e^{2\mu_{m}+\sigma_{m}^2}=\mathbb{D}\{r_{m}\}=\beta^{2}(\beta^{2}-1) \sum_{k=1}^{K}(P_{k}d_{mk}^{-\alpha})^{2}.
	\end{gather}
\end{subequations}
By conversion, $\mu_{m}$ and $\sigma_{m}^{2}$ is further expressed as
\begin{subequations} \label{character_approxLN}
	\begin{gather}
		\mu_{m}=2\ln\left(\mathbb{E}\left\{r_m\right\}\right)-\frac{1}{2}\ln\left(\mathbb{E}^{2}\{r_m\}+\mathbb{D}\{r_m\}\right),
		\\
		\sigma_{m}^{2}=\ln\left(\mathbb{E}^{2}\{r_m\}+\mathbb{D}\{r_m\}\right)-2\ln\left(\mathbb{E}\left\{r_m\right\}\right).
	\end{gather}
\end{subequations}
After approximation, $r_{m}$ becomes a LN random variable, i.e., $\widetilde{r}_{m}$ and $\ln(\widetilde{r}_{m})$ obtained is a normal random variable with an exact expression of PDF.
\begin{equation}
	\ln(\widetilde{r}_{m}) \sim \mathcal{N} (\mu_{m},\sigma_{m}^2).
\end{equation}
Since $\ln(\widetilde{r}_{m})$ has a known PDF after approximation, we will exploit the maximum likelihood estimator to solve the optimization problem proposed in the following so that the RMSL under shadow fading can be realized.

\subsection{The Maximum Likelihood Estimation Method}
Firstly, we define that $\bm{u}=[u_1, u_2,\cdots,u_K]^T$, $\bm{v}=[v_1,v_2,\cdots,v_K]^T$ and $\bm{P}=[P_1,P_2,\cdots,P_K]^T$, where $[\bm{u},\bm{v}]$ represents the locations of TSs and $\bm{P}$ denotes the power of TSs. From (\ref{Euclidean_distance}), (\ref{character_rm}) and (\ref{character_approxLN}), we can find that the mean and variance of $\ln(\widetilde{r}_{m})$, i.e., $\mu_{m}$ and $\sigma_{m}^2$, are the non-linear functions of $\bm{u}$, $\bm{v}$ and $\bm{P}$. Therefore, by defining that $\bm{\theta}=(u_{1},u_{2},\cdots,u_{K},v_{1},v_{2},\cdots,v_{K},P_{1},P_{2},\cdots,P_{K})^T$, the likelihood function can be written as
\begin{align} \label{lik_theta}
	L(\bm{\theta})
	&=
	f(\hat{y}_1,\hat{y}_2,\cdots,\hat{y}_M;\bm{\theta})\notag\\
	&=  
	\prod_{m=1}^{M}\frac{1}{\sqrt{2\pi\sigma_{m}^{2}}}e^{-\frac{(\hat{y}_m-\mu_{m})^{2}}{2\sigma_{m}^{2}}}\notag\\
	& = 
	\prod_{m=1}^{M}\frac{1}{\sqrt{2\pi\sigma_{m}^{2}}}e^{-\frac{(\ln(\hat{r}_{m})-\mu_{m})^{2}}{2\sigma_{m}^{2}}},
\end{align}
where $\hat{r}_m$ represents the observed RSS measurement, $\hat{y}_m$ is the logarithm of $\hat{r}_m$ and $f(\cdot)$ denotes the probability density function. Taking the natural logarithm of both sides yields
\begin{align} \label{ln_lik_theta}
	\ln(L(\bm{\theta}))
	&=  
	\ln\left(\prod_{m=1}^{M}\frac{1}{\sqrt{2\pi\sigma_{m}^{2}}}e^{-\frac{\left(\ln(\hat{r}_{m})-\mu_{m}\right)^{2}}{2\sigma_{m}^{2}}}\right)\notag\\
	& = 
	-\frac{1}{2}M\ln(2\pi)-\frac{1}{2}\sum_{m=1}^{M}\ln(\sigma_{m}^{2})-\sum_{m=1}^{M}\frac{\left(\ln(\hat{r}_{m})-\mu_m\right)^{2}}{2\sigma_{m}^{2}}\notag\\
	& =
	-\frac{1}{2}\sum_{m=1}^{M}\left(\ln(\sigma_{m}^{2})+\frac{\left(\ln(\hat{r}_{m})-\mu_m\right)^{2}}{\sigma_{m}^{2}}\right) + const,
\end{align}
where $const$ has nothing to do with the positions of the extreme points of the function. Hence, we need to maximize the objective function
\begin{equation} \label{ln_lik_theta1}
	-\frac{1}{2}\sum_{m=1}^{M}\left(\ln(\sigma_{m}^{2})+\frac{\left(\ln(\hat{r}_{m})-\mu_m\right)^{2}}{\sigma_{m}^{2}}\right)
\end{equation}
based on MLE criterion. For the simplicity of calculation, the objective function (\ref{ln_lik_theta1}) is reformulated as
\begin{equation} \label{Obj_Fun}
	\sum_{m=1}^{M}\left(\ln(\sigma_{m}^{2})+\frac{\left(\ln(\hat{r}_{m})-\mu_m\right)^{2}}{\sigma_{m}^{2}}\right),
\end{equation}
and we need to minimize the objective function (\ref{Obj_Fun}) via the optimization problem as follows:
\begin{subequations} \label{optimization1}
	\begin{align}
		&\mathop{\min}_{u_{k},v_{k},P_{k}}\sum_{m=1}^{M}\left(\ln\left(\sigma_{m}^{2}\right)+\frac{\left(\ln\left(\hat{r}_{m}\right)-\mu_m\right)^{2}}{\sigma_{m}^{2}}\right)
		\\
		&\text{s.t.}\quad 0 \leq u_{k} \leq l,
		\\
		&\, \qquad 0 \leq v_{k} \leq w,
		\\
		&\, \qquad P_{low} \leq P_{k} \leq P_{high},
		\\
		&\, \qquad \mu_{m}=\chi\left(u_{k},v_{k},P_{k}\right),
		\\
		&\, \qquad \sigma_{m}^{2}=\psi\left(u_{k},v_{k},P_{k}\right),
	\end{align}
\end{subequations}
where $\chi\left(u_{k},v_{k},P_{k}\right)$ and $\psi\left(u_{k},v_{k},P_{k}\right)$ are given as follows:
\begin{align}
	&\chi\left(u_{k},v_{k},P_{k}\right)\notag\\
	=&2\ln\left(\mathbb{E}\left\{r_m\right\}\right)-\frac{1}{2}\ln\left(\mathbb{E}^{2}\{r_m\}+\mathbb{D}\{r_m\}\right)\notag\\
	=&2\ln\left(\beta\sum_{k=1}^{K}P_{k}\omega\left(u_{k},v_{k}\right)^{-\alpha}\right)\notag\\
	&-\frac{1}{2}\ln\left(\beta^2\left(\sum_{k=1}^{K}P_{k}\omega\left(u_{k},v_{k}\right)^{-\alpha}\right)^2+\beta^{2}(\beta^{2}-1) \sum_{k=1}^{K}(P_{k}\omega\left(u_{k},v_{k}\right)^{-\alpha})^{2}\right),
\end{align}
\begin{align}
	&\psi\left(u_{k},v_{k},P_{k}\right)\notag\\
	=&\ln\left(\mathbb{E}^{2}\{r_m\}+\mathbb{D}\{r_m\}\right)-2\ln\left(\mathbb{E}\left\{r_m\right\}\right)\notag\\
	=&\ln\left(\beta^2\left(\sum_{k=1}^{K}P_{k}\omega\left(u_{k},v_{k}\right)^{-\alpha}\right)^2+\beta^{2}(\beta^{2}-1) \sum_{k=1}^{K}(P_{k}\omega\left(u_{k},v_{k}\right)^{-\alpha})^{2}\right)\notag\\
	&-2\ln\left(\beta\sum_{k=1}^{K}P_{k}\omega\left(u_{k},v_{k}\right)^{-\alpha}\right),
\end{align}
where $\omega\left(u_{k},v_{k}\right)$ is defined as
\begin{equation}
	\omega\left(u_{k},v_{k}\right)=\sqrt{(u_k-u_m)^2+(v_k-v_m)^2}.
\end{equation}
This optimization problem with linear constraints (\ref{optimization1}) can be widely solved by Interior-Point method, Gradient Projection (GP) method, Sequential Quadratic Programming (SQP) method and Trust Region Reflective method based on gradient descent. Here the GP method is provided to solve (\ref{optimization1}) as follows:

First, we define that
\begin{equation}
	\bm{A}=\left(                 
	\begin{array}{c}   
		\bm{I}_{3K}\\  
		\bm{-I}_{3K}\\  
	\end{array}
	\right),\quad
    \bm{x}=\bm{\theta},\quad
    \bm{b}=\left(                 
    \begin{array}{c}   
    	0\cdot\bm{1}_{2K}\\
    	P_{low}\cdot\bm{1}_{K}\\
    	-l\cdot\bm{1}_{K}\\
    	-w\cdot\bm{1}_{K}\\
    	-P_{high}\cdot\bm{1}_{K}\\
    \end{array} \right).
\end{equation}
Then, the linear constraint of the optimization problem (\ref{optimization1}) can be expressed as $\bm{A}\bm{x} \geq \bm{b}$.
\begin{itemize}
	\item[step1]
	Given an initial feasible point $\bm{x_0} \in \mathbb{R}^{3K}$ and let $k:=0$.
	\item[step2]
	Determine the efficient constraint $\bm{A_1}\bm{x_k}=\bm{b_1}$ and the ineffective constraint $\bm{A_2}\bm{x_k} > \bm{b_2}$, where
	\begin{equation}\label{step2}       
	\bm{A}=\left(                 
	\begin{array}{c}   
		\bm{A_1}\\  
		\bm{A_2}\\  
	\end{array}
	\right), \quad               
	\bm{b}=\left(                 
	\begin{array}{c}   
		\bm{b_1}\\  
		\bm{b_2}\\  
	\end{array}
	\right).                 
\end{equation}
    \item[step3]
    Let $\bm{M}=\bm{A_1}$. If $\bm{M}=\varnothing$, then let $\bm{P}=\bm{I}$; otherwise, let
    \begin{equation}\label{step3}
    	\bm{P}=\bm{I}-\bm{M}^T\left(\bm{M}\bm{M}^T\right)^{-1}\bm{M}.
    \end{equation}
    \item[step4]
    Calculate
    \begin{equation}\label{step4}
    	\bm{d}_k=-\bm{P}\nabla f(\bm{x}_k).
    \end{equation}
    If $\left\|\bm{d}_k\right\| \neq 0$, go to step6; otherwise, go to step5.
    \item[step5]
    Calculate
    \begin{equation}\label{step5}
    	\bm{\omega}=\left(\bm{M}\bm{M}^T\right)^{-1}\bm{M}\nabla f(\bm{x}_k)=
    	\left(                 
    	\begin{array}{c}   
    		\bm{\lambda}\\  
    		\bm{\mu}\\  
    	\end{array}
    	\right).
    \end{equation}
    If $\bm{\lambda} \geq \bm{0}$, stop calculation and output $\bm{x}_k$ as the KKT point; otherwise, select a negative component of $\bm{\lambda}$, such as $\lambda_j < 0$, and correct the matrix $\bm{A_1}$, i.e., remove the row corresponding to $\lambda_j$ in $\bm{A}_1$, then return to step3.
    \item[step6]
    Solve the following inexact line search problem based on Armijo rule to determine the step factor $\alpha_k$:
    \begin{subequations}\label{linesearch}
    		\begin{align}
    		&\mathop{\min} f(\bm{x}_k+\alpha \bm{d}_k),
    		\\
    		&\text{s.t.}\quad 0 \leq \alpha \leq \bar{\alpha},
    	\end{align}
    \end{subequations}
    
where $\bar{\alpha}$ is determined by
    \begin{equation}
    	\bar{\alpha} =  
    	\begin{cases}
    		\mathop{\min}\left\{\frac{\left(\bm{b}_2-\bm{A}_2\bm{x}_k\right)_i}{\left(\bm{A}_2\bm{d}_k\right)_i}\mid\left(\bm{A}_2\bm{d}_k\right)_i < 0\right\}, & \bm{A}_2\bm{d}_k \ngeq \bm{0},\\
    		+\infty, & \bm{A}_2\bm{d}_k \geq \bm{0}.
    	\end{cases}	
    \end{equation}
    \item[step7]
    Let $\bm{x}_{k+1}:=\bm{x}_k+\alpha_k\bm{d}_k$, $k:=k+1$, and return to step2.
\end{itemize}

However, unlike convex optimization problems, the local optimal solution is also the global optimal solution, (\ref{optimization1}) is a non-convex and non-linear optimization problem, which has multiple local optimal solutions. Consequently, it is also intractable to find the global optimal solution via gradient descent methods. Then, a novel initial-point-determination method based on SR-WAC is proposed to approximate a better local optimal solution of (\ref{optimization1}) in the section IV against the property that the gradient descent method is sensitive to the selection of the initial value.

\section{Sparse Recovery and Weighted Average \\of Candidates-Based Method}
In this section, we are devoted to find a suitable initial value for the aforementioned non-convex and non-linear optimization problem (\ref{optimization1}) due to its multiple local optimal solutions. Hence, an efficient algorithm based on sparse recovery and weighted average of candidates (SR-WAC) is put forward to provide a better solution for (\ref{optimization1}) and improve the performance of RMSL. In detail, the SR-WAC algorithm is divided into two stages: sparse recovery exploiting the BPDN algorithm and weighted average based on k-means clustering and averaging rule. At last, the proposed maximum likelihood estimation algorithm based on SR-WAC initial-point determination is presented in its entirety and the computational complexity is given.
\subsection{Sparse Recovery Exploiting the BPDN Algorithm}
Motivated by compressive sensing (CS) theory, we roughly estimate the locations of TSs by sparse representation as the first stage. Firstly, let us consider that the ROI is discretized into a set of rectangular grids $\mathcal G=\left\{\left(u_n, v_n\right), n=1,\cdots,N\right\}$, where the length and width between GPs are $l_{grid}=\frac{l}{\sqrt{N}-1}$ and $w_{grid}=\frac{w}{\sqrt{N}-1}$, respectively. Ignoring the shadow fading, the RSS at the $m$th SN can be written as
\begin{equation}
	r_m=\sum_{k=1}^{K}P_k d_{mk}^{-\alpha}.
\end{equation}
Then, we assume that the TSs are located at the GPs of $\mathcal G$ so that $\bm{s}$ is introduced to describe whether some GP carries the TS. Hence, the RSS without obstacles received by M SNs in theory can be further written as
\begin{equation}
	\bm{r}=\bm{\Phi}\bm{s},
\end{equation}
where $\bm{\Phi}$ denotes energy propagation matrix between SNs and GPs, i.e., $[\bm{\Phi}]_{m, n}=P_{n}d_{mn}^{-\alpha}$. $d_{mn}$ represents the Euclidean distance, $P_n=P_{high}$ and $s_{n} \in [0,1]$. As only K of the N GPs carry TSs and $N \gg K$, $\bm{s}$ is K-sparse. As a result, we consider expressing the RMSL  problem as a linear sparse coding problem as follows.
\begin{subequations} \label{optimization2}
	\begin{align}
		&\mathop{\min}_{\bm{s}}\left\|\hat{\bm{r}}-\bm{\Phi}\bm{s}\right\|_2
		\\
		&\text{s.t.}\quad \left\|\bm{s}\right\|_0 \leq K,
		\\
		&\, \qquad s_n \in [0,1].
	\end{align}
\end{subequations}
As mentioned above, $\hat{\bm{r}}$ is the real observed RSS measurement.
Unfortunately, such a non-convex problem with $l_0$-norm is NP-hard. Instead, we utilize Basis Pursuit De-Noising (BPDN) algorithm in CS to reconstruct sparse signals, which approximate $l_0$-norm by $l_1$-norm. Then, the optimization problem (\ref{optimization2}) is relaxed as
\begin{subequations} \label{optimization3}
	\begin{align}
		&\mathop{\min}_{\bm{s}}\frac{1}{2}\left\|\hat{\bm{r}}-\bm{\Phi}\bm{s}\right\|_2^{2}+\lambda\left\|\bm{s}\right\|_1
		\\
		&\;\text{s.t.}\quad 0 \leq s_n \leq 1.
	\end{align}
\end{subequations}
The above problem (\ref{optimization3}) can be transformed into a quadratic programming problem and the derivation is given as follows:
\begin{subequations}
	\begin{align}
		&\frac{1}{2}\left\|\hat{\bm{r}}-\bm{\Phi}\bm{s}\right\|_2^{2}+\lambda\left\|\bm{s}\right\|_1
		\\
		=&\frac{1}{2}\left(\hat{\bm{r}}^{T}\hat{\bm{r}}-\hat{\bm{r}}^{T}\bm{\Phi}\bm{s}-\bm{s}^{T}\bm{\Phi}^{T}\hat{\bm{r}}+\bm{s}^{T}\bm{\Phi}^{T}\bm{\Phi}\bm{s}\right)+\lambda\bm{1}^{T}_{N}\bm{s}
		\\
		=&\frac{1}{2}\left(\hat{\bm{r}}^{T}\hat{\bm{r}}-2\bm{b}^{T}\bm{s}+\bm{s}^{T}\bm{B}\bm{s}\right)+\lambda\bm{1}^{T}_{N}\bm{s}
		\\
		=&\frac{1}{2}\hat{\bm{r}}^{T}\hat{\bm{r}}+\bm{c}^{T}\bm{s}+\frac{1}{2}\bm{s}^{T}\bm{B}\bm{s},
	\end{align}
\end{subequations}
where $\bm{b}=\bm{\Phi}^{T}\hat{\bm{r}}$, $\bm{c}=\lambda\cdot\bm{1}_{N}-\bm{b}$, $B=\bm{\Phi}^{T}\bm{\Phi}$. Since $\frac{1}{2}\hat{\bm{r}}^{T}\hat{\bm{r}}=const$, it can be neglected. Consequently, (\ref{optimization3}) can be evolved into
\begin{subequations} \label{optimization4}
	\begin{align}
		&\mathop{\min}_{\bm{s}}\frac{1}{2}\bm{s}^{T}\bm{B}\bm{s}+\bm{c}^{T}\bm{s}
		\\
		&\text{s.t.}\quad 0 \leq s_n \leq 1, \quad n=1,2,\cdots,N.
	\end{align}
\end{subequations}
(\ref{optimization4}) is a quadratic programming problem with a standard form and can be solved effectively by many basic algorithms such as interior-point method, active-set method and so on.
\begin{figure}[t]
	\centering
	\epsfig{figure=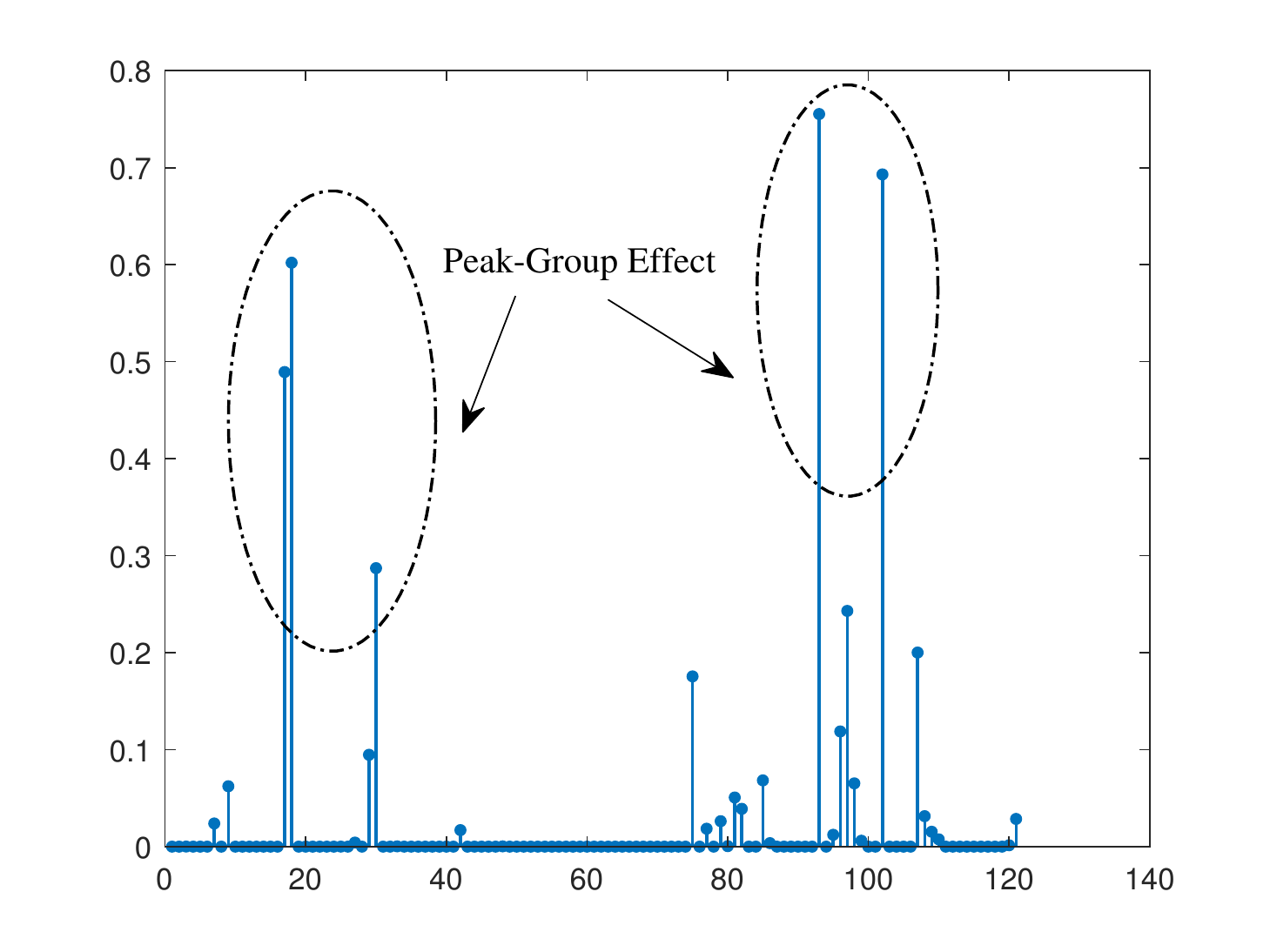,width=8cm}
	\caption{Sparse recovery based on BPDN algorithm}
	\label{fig_SparseRecovery}
\end{figure}
As shown in Fig. \ref{fig_SparseRecovery}, $\bm{s}^*$ obtained from solving (\ref{optimization4}) is not necessarily K-sparse owing to the fact that the TSs are almost off-grid. Furthermore, the power of TSs is mainly distributed into the nearby GPs, which is called Peak-Group Effect. Then, the weighted average of candidates is considered to achieve the approximate locations of the TSs as the second stage.

\subsection{Weighted Average of Candidates Based on K-means Clustering and Averaging Rule}

\begin{figure*}[t]
	\centering
	\subfigure[Adaptive dynamic threshold truncation]
	{\includegraphics[width=8cm,height=7.5cm]{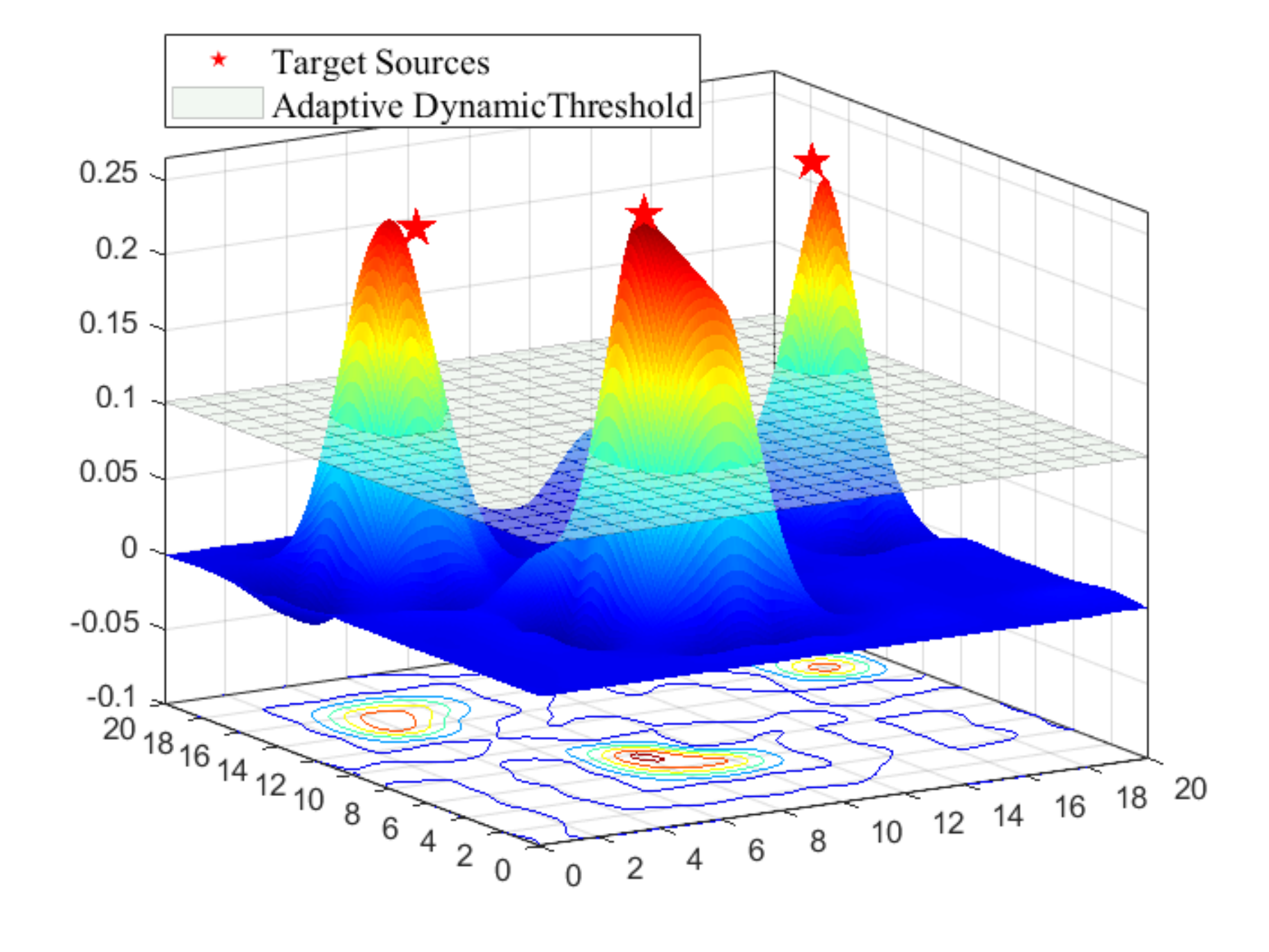}} \hspace{1cm}
	\subfigure[K-means clustering and weighted avarage]
	{\includegraphics[width=8cm,height=7.5cm]{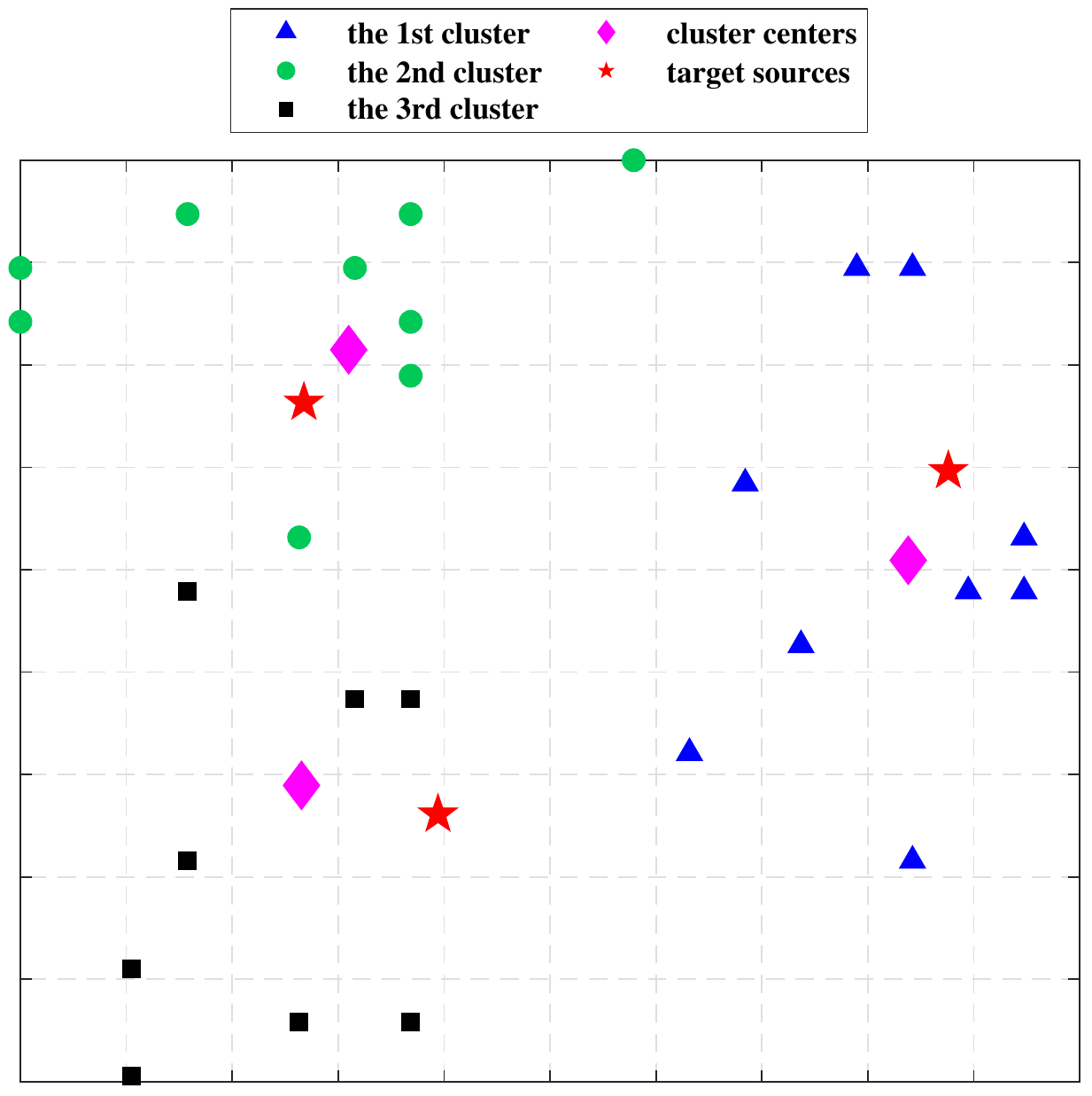}}
	\caption{weighted average of candidates based on k-means clustering and averaging rule.}
	\label{weighted average of candidates}
\end{figure*}

To cluster the GPs, We firstly define an adaptive dynamic threshold (ADT)
\begin{equation} \label{Truncation}
	Thr=max(\bm{s}^{*})-std(\bm{s}^{*}),
\end{equation}
where $std(\cdot)$ represents the standard deviation of the vector. Via the ADT truncation, the GPs near the TSs are selected to the most extent and the GPs whose power is below ADT are abandoned. The advantage of ADT is that when the noise level is high, ADT will lower the threshold against noise interference, ensuring more GPs nearby the TSs are reserved in the ROI and on the contrary, ADT will raise the threshold when the noise level is low so as to reduce the number of selected GPs weakly correlated with the TSs. Fig. \ref{weighted average of candidates}(a) shows 3-D adaptive dynamic threshold truncation.

Next, k-means clustering algorithm is considered to divide the candidates set $\mathcal D$ selected via ADT truncation into $K$ clusters, denoted as $\Pi_1$, $\Pi_2$, $\cdots$, $\Pi_K$. Then, by weighting the grid points in $\mathcal D$ with the averaging rule, cluster centers, i.e., the estimated TSs roughly, are obtained, written as:
\begin{subequations} \label{average_rule}
	\begin{gather}
		u_k^{0}=\frac{\sum_{n \in \Pi_k}s_n^{*}u_n}{\sum_{n \in \Pi_k}s_n^{*}},
		\\
		v_k^{0}=\frac{\sum_{n \in \Pi_k}s_n^{*}v_n}{\sum_{n \in \Pi_k}s_n^{*}}.
	\end{gather}
\end{subequations}
Moreover, we define the power of estimated TSs as
\begin{equation}\label{power_establishment}
	P_k^0 = \max\limits_{n \in \Pi_k}s_n^{*}P_{high}.
\end{equation}
Using (\ref{average_rule}) and (\ref{power_establishment}), initial estimated TSs set
\begin{equation} \label{Initial_point}
	\mathcal T = \left\{\left(u_1^{0}, v_1^{0}, P_1^{0}\right),\left(u_2^{0}, v_2^{0}, P_2^{0}\right), \cdots, \left(u_K^{0}, v_K^{0}, P_K^{0}\right)\right\}
\end{equation} 
is obtained as the initial point of the optimization problem (\ref{optimization1}). Fig. \ref{weighted average of candidates}(b) shows the result of clustering and weighted average in the case of three TSs.

\subsection{The Proposed MLE Algorithm Progress and Computational Complexity Analysis}
Based on the above analysis, the proposed Maximum Likelihood Estimation (MLE) algorithm based on sparse recovery and weighted average of candidates (SR-WAC) is summarized in Algorithm \ref{Alg_MLE}.
\begin{algorithm}[t]
	\begin{small}
		\caption{Maximum Likelihood Estimation Based on Sparse Recovery and Weighted Average of Candidates}
		\label{Alg_MLE}
		\KwIn{$\bm{\Phi}$, $\mathcal G$, $\bm{a}$, $\bm{\hat{r}}$}
		\KwOut{ $\Omega$\;}
		Initialize $K$, $N$, $M$, $\alpha$, $\sigma_{s}$, $P_{high}$, $P_{low}$\;
		\tcp{Sparse recovery exploiting BPDN}
		$\lambda = 10^{-3}$\; 
		Calculate $\bm{b}$, $\bm{c}$, $\bm{B}$ using $\mathcal G$, $\bm{\Phi}$, $\bm{\hat{r}}$, $\lambda$\;
		Calculate $\bm{s}^{*}$ by solving (\ref{optimization4})\;
		\tcp{Adaptive dynamic threshold truncation}
		Calculate $Thr$ using (\ref{Truncation})\;
		Obtain $\mathcal D$ by using $Thr$ to truncate $\bm{s}^{*}$\;
		\tcp{Weighted average}
		Obtain $K$ clusters $\Pi_1$, $\cdots$, $\Pi_K$ using k-means clustering,\\
		Obtain $\mathcal T$ using (\ref{average_rule}) and (\ref{power_establishment})\;
		\tcp{Maximum likelihood estimation}
		Construct the $\bm{x_0}$ by utilizing $\mathcal T$ in the form of $\bm{\theta}$ and let $k:=0$\;
		\While{convergence criterion not met }{
		Obtain $\bm{A}_1$, $\bm{A}_2$, $\bm{b}_1$, $\bm{b}_2$ by (\ref{step2})\;
		Let $\bm{M}=\bm{A}_1$\;
		\uIf{$\bm{M}=\varnothing$}{Let $\bm{P}=\bm{I}$\;}
        \Else{Calculate $\bm{P}$ using (\ref{step3})\;}
        Calculate $\bm{d}_k$ by (\ref{step4})\;
        \uIf{$\left\|\bm{d}_k\right\| \neq 0$}{
        Calculate $\alpha_k$ by solving (\ref{linesearch})\;
        Let $\bm{x}_{k+1}:=\bm{x}_k+\alpha_k\bm{d}_k$ and $k:=k+1$\;
        Return to $\bm{11}$\;}
        \Else{Calculate $\bm{\omega}$, $\bm{\lambda}$, $\bm{\mu}$ using (\ref{step5})\;
        \uIf{$\bm{\lambda} \geq \bm{0}$}{break\;}
        \Else{Correct $\bm{A}_1$ and return to $\bm{12}$}
        }
        }
        \KwRet{$\bm{x}_{k}$}\;
		Calculate $\Omega=\left\{\left(u_{k}^{*}, v_{k}^{*}, P_{k}^{*}\right), k=1, \cdots, K\right\}$ by reconstructing $\bm{x}_{k}$;
	\end{small}	
\end{algorithm}

Before analyzing the computational complexity, we firstly assume that the number of elements in the candidate set $\mathcal D$ is $H$ and $K < H \ll M \approx N$. Then, the complexity is given as follows. In the sparse recovery, since the original problem (\ref{optimization2}) is reformulated as (\ref{optimization4}) exploiting the BPDN algorithm, which is a quadratic programming problem. So the cost can be computed as $O(N^3)$. Then, in the K-means clustering, it has a complexity of $O(DHKI)$, where $D$ and $I$ denote the dimension of the data set and the iteration number, respectively. Thus, the cost in this stage is simplified as $O(KH)$. Finally, we utilize the GP method to solve the optimization problem (\ref{optimization1}) and this method is used to find a local optimal solution after a number of iterations, $I_{GP}$. Consequently, the cost of GP method can be written as $O(I_{GP}K^3)$, simplified as $O(K^3)$. To sum up, the whole algorithm has a complexity of $O(N^3+KH+K^3)$, which yields $O(N^3)$, i.e., a cubic order of complexity in the number of grid points. Thus, the proposed method has a lower computational complexity compared with the MMSE algorithm, whose complexity is $O(K^3N^3)$.

\section{Numerical Simulations}
In this section, we are devoted to evaluating the performance of the proposed MLE method based on SR-WAC for RMSL by numerical simulations and compare it with the other method in \cite{Zandi2019Multi}.

In the simulation setup, a 2000m by 2000m area is considered. M sensor nodes and $K=3$ source nodes are randomly distributed in the ROI, where M is drawn from $\left[60, 180\right]$. The path-loss exponent (PLE) $\alpha=2.5$ is assumed and the shadow fading factor $\sigma$ is drawn from $\left[2, 10\right]$ dB which are typical values for practical channels, depending on the severity of the shadow fading. Moreover, the transmitted power $P_k$ of three source nodes is randomly drawn from $\left[2000, 4000\right]$ mW.

The evaluation metrics are root-mean-square error (RMSE), maximum error function (MEF) and cumulative distribution function (CDF) of the average error, where the definitions of RMSE and MEF are given as follows. RMSE is defined as
\begin{equation}
	RMSE =\frac{1}{J} \sum_{j=1}^{J}\sqrt{\frac{1}{K}\sum_{k=1}^{K}\left(\left(u_{jk}-u_{jk}^*\right)^2+\left(v_{jk}-v_{jk}^*\right)^2\right)},
\end{equation}
where $(u_{jk},v_{jk})$ and $(u_{jk}^*,v_{jk}^*)$ are the location coordinates of source nodes and estimated nodes in the $j$th realization, respectively. The maximum error of $K$ source nodes for every realization is defined as:
\begin{equation}
	\Delta = \max\limits_{k=1, \cdots, K}\sqrt{\left(u_k-u_k^*\right)^2+\left(v_k-v_k^*\right)^2}.
\end{equation}
Thus, MEF is written as
\begin{equation}
	MEF=P_r(\Delta > \emph{d}) =1-F_{\Delta}(\emph{d}),
\end{equation}
indicating the probability that the positioning error of at least one source node is greater than $d$, where $F_{\Delta}(\emph{d})$ represents the CDF of $\Delta$.

The MEF underlines the influence of the worst positioning result in each localization on the positioning performance, while the CDF of the average error makes full use of all positioning results in each localization to evaluate the positioning performance.

The simulation results are the output of $J=5000$ random trials carried out in MATLAB R2020a and we examine the performance of different situations presented as follows.

\subsection{Effects of Shadow Fading Strength}
In this simulation, we compare the performance of the proposed method with the MMSE in \cite{Zandi2019Multi}. First, we set $M=150$, $N=121$ as fixed parameters.
\begin{table}[t]
	\newcommand{\tabincell}[2]{\begin{tabular}{@{}#1@{}}#2\end{tabular}}
	\centering
	\caption{Relative RMSE of location estimation versus $\sigma$}
	\begin{tabular}{ c c c c c c  }
	\toprule
	\bm{$\sigma$}[dB]   &  \tabincell{c}{2} & \tabincell{c}{4} & \tabincell{c}{6} &\tabincell{c}{8}  & \tabincell{c}{10}\\
	\midrule
	\textbf{MMSE}  	    &0.1911     &0.2209     &0.2474    &0.2719    &0.2923\\
	\textbf{ML}      &0.0520     &0.0716     &0.1016   &0.1396    &0.1655\\
	\bottomrule	
\end{tabular}
\label{tab:RMSE_sigma}
\end{table}

\begin{figure*}[t]
	\centering
	\subfigure[]
	{\includegraphics[width=8cm,height=7.5cm]{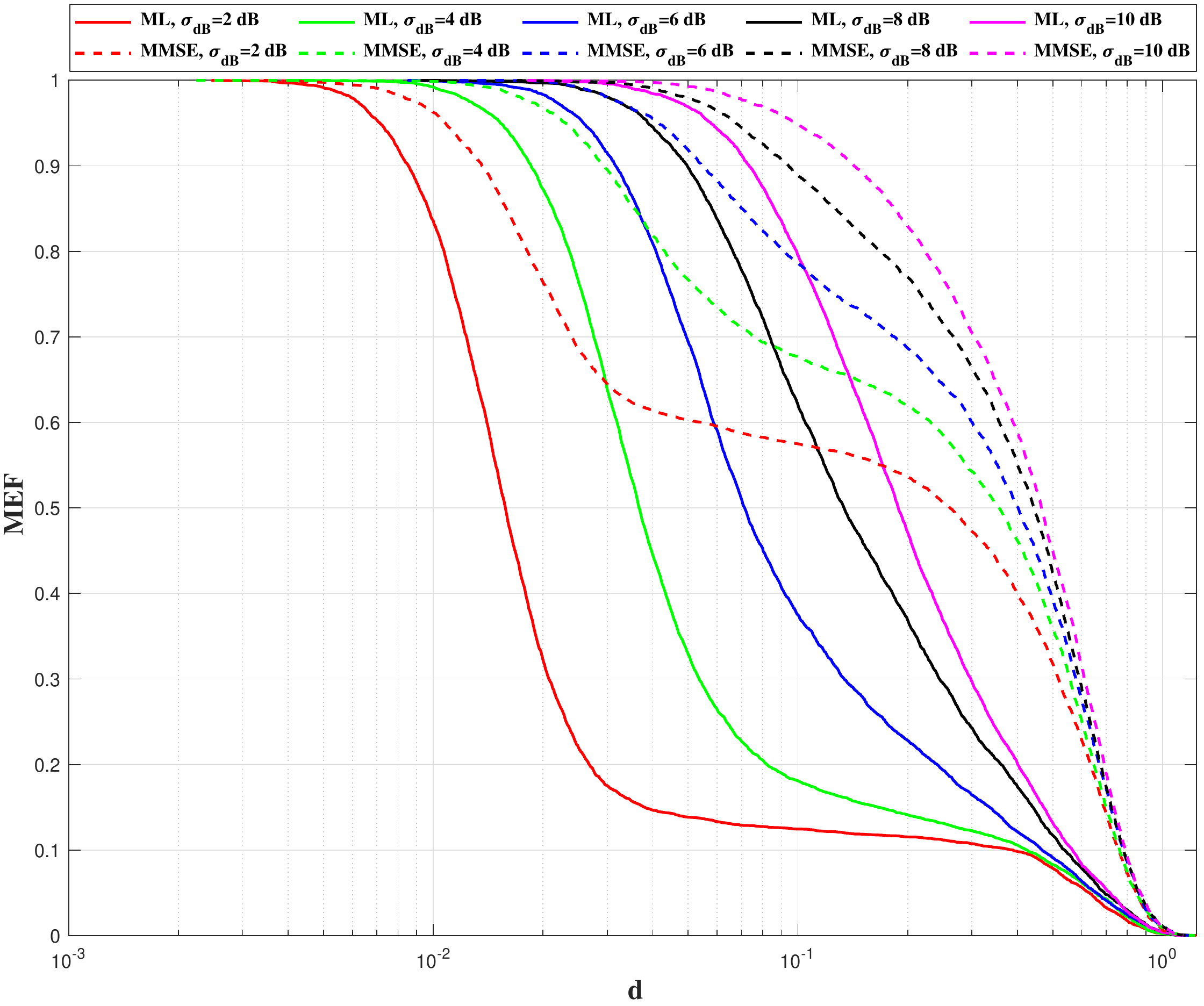}} \hspace{1cm}
	\subfigure[]
	{\includegraphics[width=8cm,height=7.5cm]{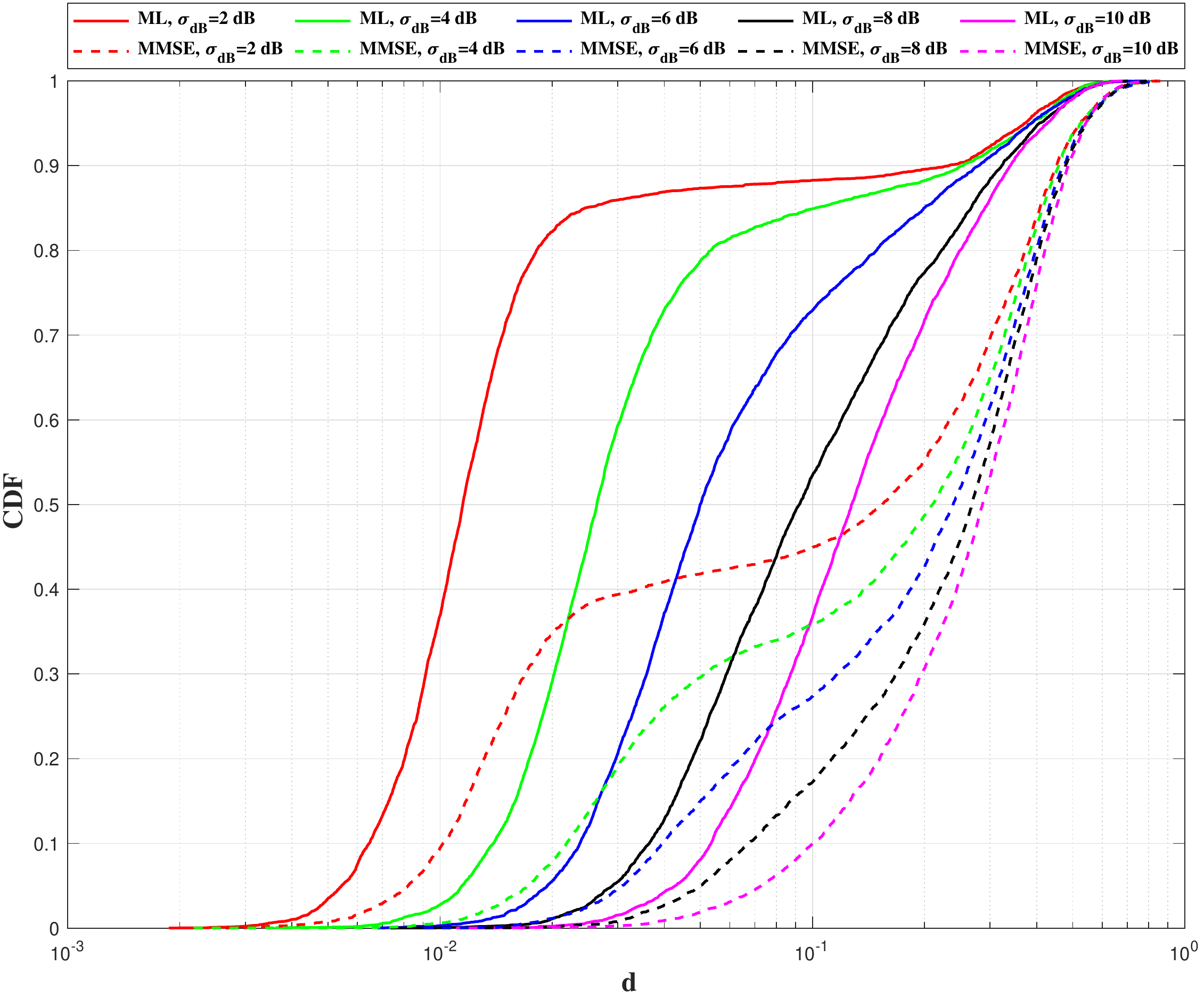}}
	\caption{The performance metrics of different approaches versus \bm{$\sigma$} (dB) for (a) MEF and (b) CDF.}
	\label{shadowing}
\end{figure*}
Table. \ref{tab:RMSE_sigma} presents the relative RMSE results versus the shadow fading strength, which is given by
\begin{equation}
	relative\ RMSE = \frac{RMSE}{\sqrt{ROI's\quad area}},
\end{equation} 
when shadow fading factor $\sigma$ is changed from 2 dB to 10 dB at a step size of 2 dB. As the table shows, the relative RMSE of the two algorithms decreases (the localization accuracy increases) with the decrease of the strength of shadow fading. On the other hand, the relative RMSE of the proposed method is much lower than the MMSE algorithm in [44], which shows the localization accuracy of the proposed method is much better.

Fig. \ref{shadowing}(a) illustrates the MEF performance versus the maximum relative positioning error distance $\Delta$ for different values of shadow fading strength, where $\Delta$ is specifically expressed as the ratio between the maximum relative positioning error distance and the square root of the area. It is observed from the figure that the probability of maximum relative positioning error $\Delta$ for every trial greater than $d$ increases continually with the increase of shadow fading strength. Particularly for the proposed method, the probability of maximum relative positioning error $\Delta$ greater than $10\%$ is down to $12.46\%$ when $\sigma=2$dB. However, for the MMSE method, the probability reaches $57.44\%$ in the same situation. Finally, comparing two sets of curves, the proposed method has a better localization accuracy from the figure that the curves of the proposed method is at the bottom left of the other. The MEF indicates that the proposed method will greatly reduce the appearance of extreme positioning results when shadow fading strength increases.

The CDF performance for different values of shadow fading strength is shown in the fig. \ref{shadowing}(b), which utilizes the average relative positioning error to measure the positioning performance. It can be seen that like fig. \ref{shadowing}(a), the probability of average relative positioning error less than $d$ increases with the decrease of shadow fading strength. Especially for the proposed method, the probability of average relative positioning error less than $10\%$ reaches $88.28\%$ when $\sigma=2$ dB. Nevertheless, for the MMSE method, the probability is only $45.04\%$ in this case. Moreover, we can observe that the probability of average relative positioning error less than $10\%$ for the proposed method when $\sigma=10$ dB and the probability for the MMSE method when $\sigma=4$ dB are basically equal, i.e., $35.82\%$, which implies that the proposed method has a stronger anti-fading ability. The CDF indicates that the proposed method has a higher localization accuracy overall.

This simulation emphasizes the superiority of the proposed method under different shadow fading, which greatly improves the robustness of the localization against the obstacles.
\subsection{Influences of Sensor Density}
Then, beyond question, it is of great significance to consider the influence of the sensor density on RMSL. To the best of our knowledge, the deployment of more sensors will improve the localization accuracy to some extent owing to less unknown observations. In this simulation, $\sigma=6$ dB and $N=121$ are set. In addition, the number of the sensors is drawn from 60 to 180 at a step size of 30. Here, we define the sensor density, $\rho$ as:
\begin{equation}
	\rho = \frac{sensor\ numbers}{ROI's\quad area}.
\end{equation}
Therefore, we can obtain the sensor density which varies from $1.5\times 10^{-5}m^{-2}$ to $4.5\times 10^{-5}m^{-2}$ at a step size of $0.75\times 10^{-5}m^{-2}$.

\begin{table}[t]
	\newcommand{\tabincell}[2]{\begin{tabular}{@{}#1@{}}#2\end{tabular}}
	\centering
	\caption{Relative RMSE of location estimation versus $\rho$}
	\begin{tabular}{ c c c c c c}
		\toprule
		\bm{$\rho$}[$\times 10^{-5}m^{-2}$]   &  \tabincell{c}{1.5} & \tabincell{c}{2.25} & \tabincell{c}{3} &\tabincell{c}{3.75}  & \tabincell{c}{4.5}\\
		\midrule
		\textbf{MMSE}  	    &0.1911     &0.1807     &0.1777    &0.1664    &0.1633\\
		\textbf{ML}      &0.0895     &0.0706     &0.0573   &0.0503    &0.0467\\
		\bottomrule	
	\end{tabular}
	\label{tab:RMSE_sensor}
\end{table}

\begin{figure*}[t]
	\centering
	\subfigure[]
	{\includegraphics[width=8cm,height=7.5cm]{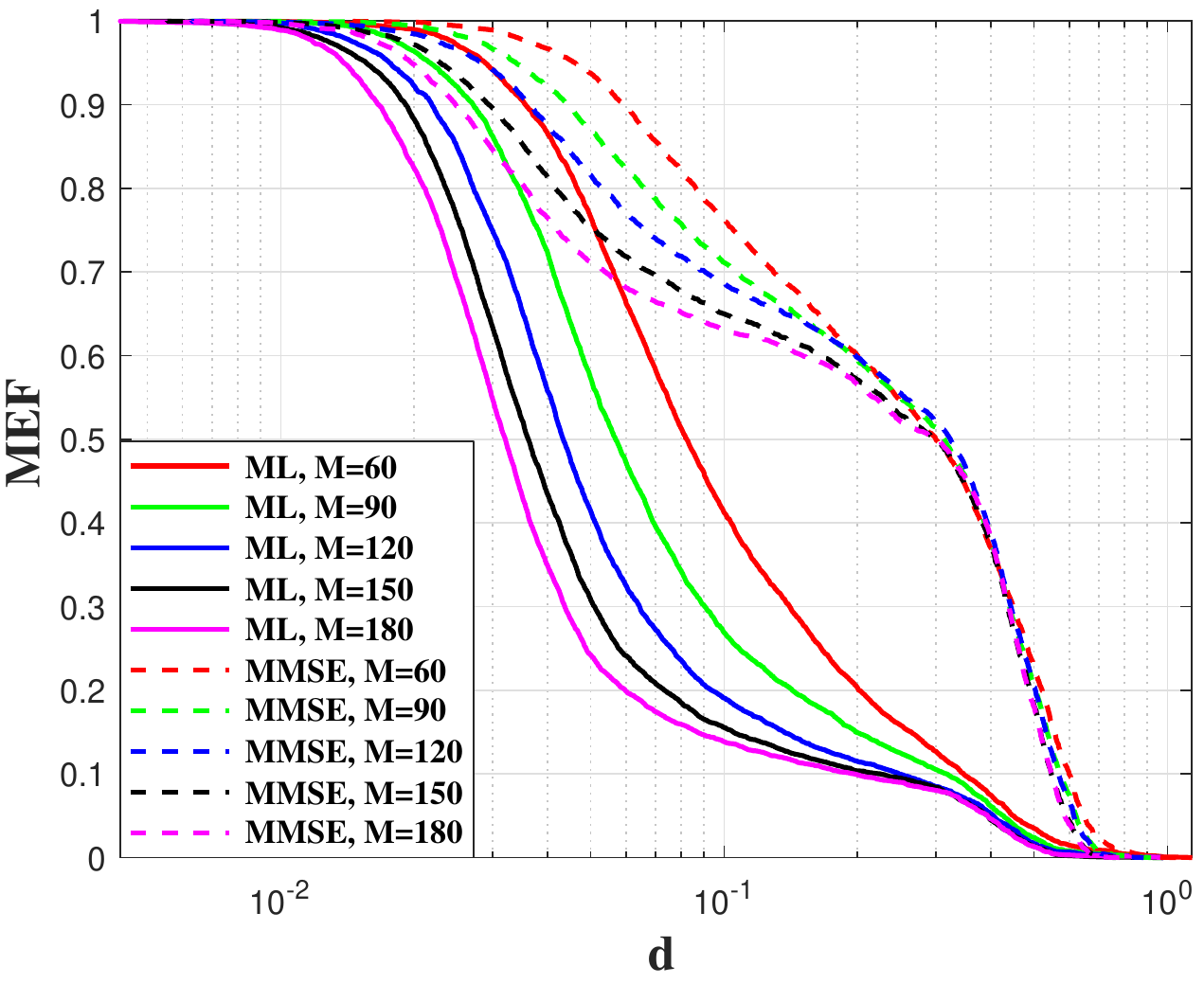}} \hspace{1cm}
	\subfigure[]
	{\includegraphics[width=8cm,height=7.5cm]{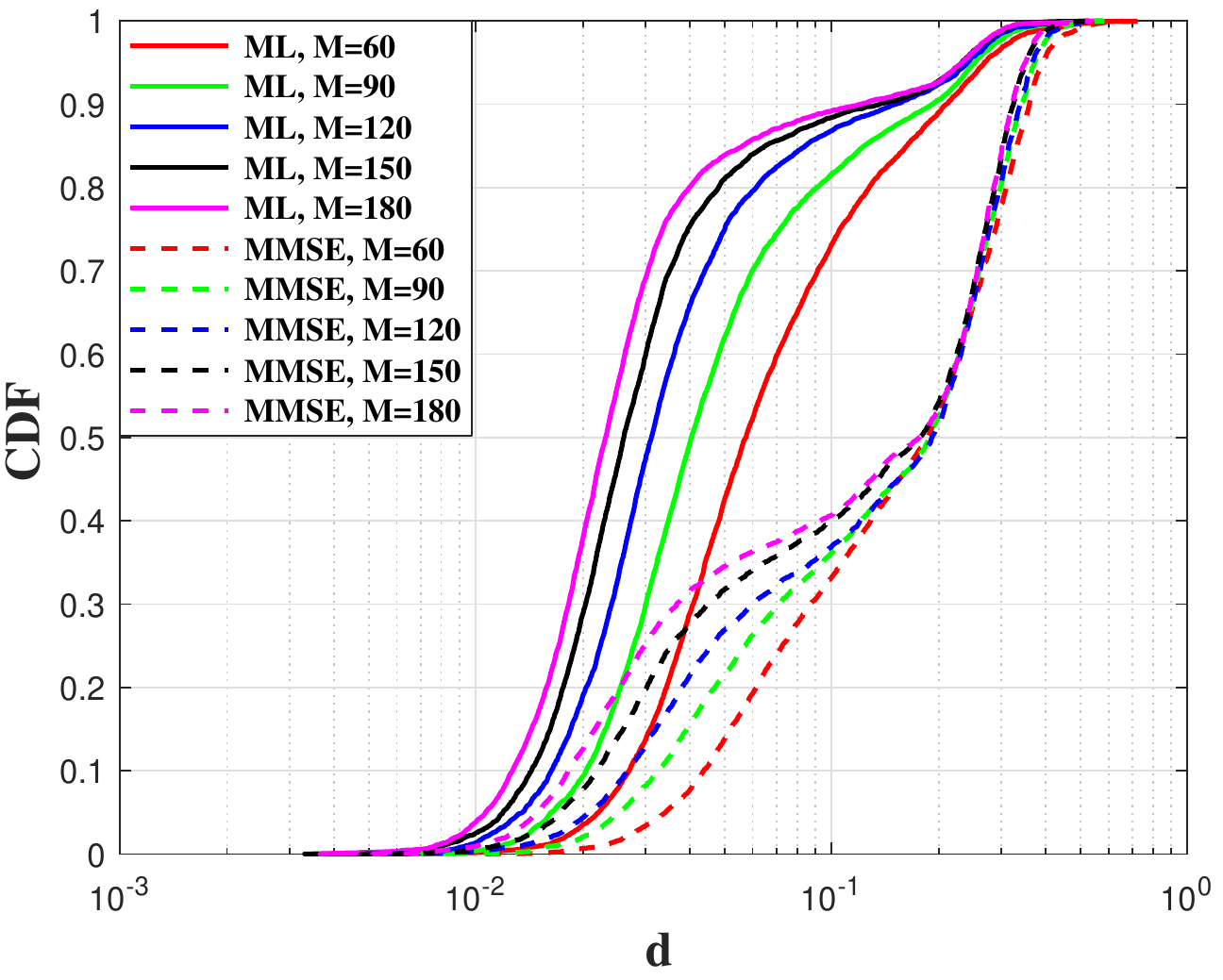}}
	\caption{The performance metrics of different approaches versus sensor numbers for (a) MEF and (b) CDF.}
	\label{sensor}
\end{figure*}

Table. \ref{tab:RMSE_sensor} illustrates the relative RMSE of the location for the two approaches under different sensor densities. As can be seen, the relative RMSE decreases, as the sensor density increases, which is consistent with our perception. At the same time, the positioning performance of the proposed approach far exceeds the other in \cite{Zandi2019Multi}. Specifically, the relative RMSE of the proposed method is all below $10\%$ and the minimum relative RMSE of the MMSE method is more than $15\%$, which indicates that the proposed approach greatly improves the localization accuracy compared with the MMSE approach.

It is observed from the MEF performance in the fig. \ref{sensor}(a) that as the sensor density increases, the probability of maximum relative positioning error more than $d$ decreases. Especially, the probability of maximum relative positioning error more than $10\%$ is down to $13.95\%$  for the proposed method when the sensor density arrives at $4.5 \times 10^{-5}m^{-2}$ ($M=180$), while the probability reaches $63.06\%$ for the MMSE method. As a result, the proposed method has a higher localization accuracy and robustness than the other in \cite{Zandi2019Multi}. 

Fig. \ref{sensor}(b) presents the CDF performance versus the average relative positioning error. As the sensor density increases, the probability of average relative positioning error less than $d$ increases. For the proposed method, the highest and lowest probabilities of average relative positioning error less than $10\%$ reach $89.30\%$ and $73.12\%$ when the sensor numbers are 180 and 60, respectively. However,for the MMSE method, the highest probability is only $40.84\%$ when the sensor number is 180. From another aspect, as the sensor density increase at the same step size, the improvement of the probability gradually slows down.

This simulation explains that the proposed method can exploit sensors more efficiently to realize a more accuracy localization than the MMSE method. On the other hand, with the increase of the sensor density when gradually approaching the maximum capacity, the positioning performance will improve slowly. Thus, in practice, how many sensors employed to realize a tradeoff between cost and performance becomes a major problem which should be considered.

\section{Conclusion}
In this paper, the multiple sources localization problem based on received-signal-strength (RSS) measurements under log-normal shadow fading is addressed. Modeled by log-normal sum (SLN), the RSS has an unknown probability density function (PDF) and classic estimators are difficult to use. We exploit the Fenton-Wilkinson (F-W) method so that the SLN variable is approximated as a LN variable, which can be analyzed with an exact PDF. Therefore, we propose a maximum likelihood estimation (MLE) based method to estimate the optimal locations of the sources. However, this optimization problem formulated is highly non-convex and mathematically intractable. For this reason, we put forward a sparse recovery and weighted average of candidates (SR-WAC) based approach to set up an initiation. Numerical simulation results shows that the MLE method based on SR-WAC provides a better localization performance compared with the approach in \cite{Zandi2019Multi} under different shadow fading strengths and sensor densities.



%

\appendices


\ifCLASSOPTIONcaptionsoff
\newpage
\fi



%
%

\bibliographystyle{IEEEtran} 
\begin{footnotesize}
	\bibliography{IEEEabrv,my_reference} 
\end{footnotesize}
\end{document}